\documentclass[lettersize,journal]{IEEEtran}
\usepackage{amsmath,amsfonts}
\usepackage{algorithmic}
\usepackage{algorithm}
\usepackage{array}
\usepackage[caption=false,font=normalsize,labelfont=sf,textfont=sf]{subfig}
\usepackage{textcomp}
\usepackage{stfloats}
\usepackage[hidelinks]{hyperref}
\usepackage{url}
\usepackage{verbatim}
\usepackage{booktabs}
\usepackage{graphicx}
\usepackage{cite}
\usepackage{siunitx}
\usepackage{orcidlink}

\usepackage{orcidlink}
\newcommand{\circled}[1]{\tikz[baseline=(char.base)]{
  \node[shape=circle,draw,inner sep=0.5pt] (char) {#1};}}
\begin{document}

\title{Data-Driven Discovery of Multiscale Power System Oscillation Governing Equations Using SINDy-SENDAI} 

\author{Andrea Pomarico\orcidlink{0009-0006-6677-6399}, Yuxuan Bao\orcidlink{0009-0006-7648-6803}, Liyao Mars Gao\orcidlink{0000-0001-9692-4264}, Salvatore Tessitore\orcidlink{0000-0002-3674-9211},\\ Giorgio Maria Giannuzzi\orcidlink{0000-0003-0314-4483}, Alberto Berizzi\orcidlink{0000-0002-2856-783X}, and J. Nathan Kutz\orcidlink{0000-0002-6004-2275}.
\thanks{A. Pomarico and A. Berizzi are with the Department of Energy, Politecnico di Milano, Milan, Italy. Y. Bao is with the Department of Applied Mathematics, University of Washington, Seattle, WA, USA. L. M. Gao is with the Department of Statistics, Stanford University, Palo Alto, CA, USA. J. N. Kutz is with Autodesk Research, London, UK. G. M. Giannuzzi and S. Tessitore are with the Department of System Engineering, Terna Rete Italia S.p.A., Rome, Italy. Corresponding author: A. Pomarico (andrea.pomarico@polimi.it).}}

\markboth{Journal of XXX}%
{Shell \MakeLowercase{\textit{et al.}}: A Sample Article Using IEEEtran.cls for IEEE Journals}


\maketitle

\begin{abstract}
Monitoring electromechanical oscillations is crucial for maintaining the stability of modern power systems, particularly in the presence of increasing penetrations of inverter-based resources (IBRs), which introduce new dynamic behaviors. In this work, we propose a hierarchical multiscale framework based on the SINDy-SENDAI algorithm to characterize the transient dynamics captured by wide-area measurements. The proposed deep learning architecture robustly separates low- and high-frequency components embedded in sensor data and incorporates a Sparse Identification of Nonlinear Dynamical Systems (SINDy) module in the latent space to identify parsimonious governing equations. In contrast to conventional deep learning approaches that often produce black-box models with limited interpretability, the proposed framework learns an explicit dynamical representation, enabling physical interpretation, stability assessment, and forecasting of electromechanical oscillations. Given the societal importance of modern power systems, the proposed approach is specifically designed to satisfy key requirements for practical deployment, namely robustness, interpretability, and stable performance under diverse operating conditions. The framework is first validated on the two-area Kundur test system using conventional modal analysis as ground truth and subsequently demonstrated on two real-world datasets: the 2016 Iberian oscillatory event and the 2021 ambient measurements from the southern Italian power grid. The results show that SINDy-SENDAI consistently outperforms the state-of-the-art Hankel-DMD method and that the learned latent dynamics are sufficiently informative to accurately reconstruct and predict the behavior of the full system in the original state space.
\end{abstract}


\section{Introduction}
\IEEEPARstart{T}{he} planning and operation of power systems have recently experienced notable changes due to the growing integration of Renewable Energy Sources (RES)~\cite{HASSAN2024100545}. In this scenario, maintaining system stability and ensuring a reliable electricity supply have become key challenges for Transmission System Operators (TSOs). The large scale and interconnectivity of modern power grids increase the probability of faults and failures, and unexpected disturbances or cascading events can trigger widespread blackouts \cite{power_system_modeling_control_2013}. Such incidents not only damage the power infrastructure but also carry serious social and economic repercussions~\cite{ENTSOE2025IberianBlackout}. To mitigate these risks, Phasor Measurement Units (PMUs) have been increasingly deployed within Wide-Area Measurement Systems (WAMS), providing valuable real-time information extracted from the streams of data.

Given the crucial role of power systems in modern society, substantial efforts have been devoted to developing advanced, data-driven tools~\cite{kutz2026data} capable of accurately capturing their complex dynamic behavior across multiple time scales. In particular, maintaining power system stability, especially with the increasing penetration of inverter-based resources (IBRs)~\cite{hatziargyriou2020definition}, has motivated the development of algorithms for real-time identification of electromechanical oscillations using PMU data. Such oscillations, if not promptly detected and properly damped, can severely compromise system stability~\cite{ENTSOE2025IberianBlackout}. Interarea oscillations are of particular concern for TSOs because they involve large power exchanges between interconnected regions and, when poorly damped, can jeopardize overall system stability~\cite{kundur2007power}. Their criticality has been demonstrated by major disturbance events, including the 2025 Iberian blackout~\cite{BOIberico}. These oscillations typically occur at frequencies between 0.1 and 0.7~Hz and are characterized by the coherent motion of large groups of generators~\cite{marconato2002electric}.

To address the modern challenge of maintaining the stability and robustness of power systems, data-driven algorithms must accommodate a diverse range of timescales and noise present in sensor measurements. Further, interpretability is critical in order for operators to understanding the underlying physical operation.  The algorithm proposed here leverages two key developments.  The {\em Sparse-measurement, EfficieNt Data AssImilation} (SENDAI) model~\cite{zhang2026sendai} is specifically designed to address the robust, multiscale reconstruction problem.  Specifically, SENDAI extracts the dominant timescales by a recursive, hiearchical extraction scheme.  This allows for the isolation of the underlying physical processes at each of their respective timescales.  The {\em Sparse Identification of Nonlinear Dynamics} (SINDy)~\cite{brunton2016discovering} algorithm is then used to discover parsimonious physics models~\cite{kutz2022parsimony} at each respective scale, thus enabling a data-driven architecture for learning interpretable physics models.  Our proposed SINDy-SENDAI algorithm is thus an efficient algorithm to go directly from data to interpretable and multiscale physics models.  As is demonstrated, SINDy-SENDAI is a highly robust and effective algorithm that is demonstrated directly on European power system.

\subsection{Bibliography Review and Goal}\label{sec:biblio}
In recent years, the widespread deployment of PMUs has prompted researchers and TSOs to explore advanced mathematical frameworks for the real-time identification of inter-area oscillations. Conventional techniques, such as the Eigensystem Realization Algorithm~\cite{6993769} and the Matrix Pencil method, have proven effective in analyzing data streams from transient events. Furthermore, alternative methodologies, including the Estimation of Signal Parameters via Rotational Invariance Technique (ESPRIT)~\cite{samal,8678764}, the Tufts–Kumaresan method~\cite{GIANNUZZI20151147}, and the Yule-Walker approach~\cite{1270937}, have been proposed for implementation in control rooms. However, several of these methods struggle to isolate interarea modes under noisy ambient conditions. To mitigate these limitations, Dynamic Mode Decomposition (DMD)~\cite{kutz2016dynamic} was introduced in~\cite{barocio2014dynamic} to represent nonlinear dynamics via locally linear operators. Notably, Hankel-DMD (hDMD) has demonstrated high efficacy in processing WAMS data for oscillation monitoring within the European power grid~\cite{vicario2022practical}.


The aforementioned approaches are primarily rooted in classical signal processing, where measurements are processed through deterministic algorithms to estimate dominant modes. Recently, however, the rapid advancement of Machine Learning (ML) has led to its extensive adoption across various engineering domains. In the context of power systems, ML has been widely investigated for diverse applications~\cite{alimi2020review}, with a particularly active research thrust focused on monitoring interarea oscillations within the WAMS framework. For instance, several studies~\cite{azman2020unified,satheesh2022identification,muhammed2023deep} have employed ML architectures to predict oscillatory behavior. Beyond prediction, data-driven strategies have been proposed for the identification and classification of wideband oscillations~\cite{gao2025data}, while ML-based schemes have been developed to support remedial actions in Wide Area Damping Control~\cite{naderi2022machine}. Notably, a SINDy framework was proposed in~\cite{cai2022online} for the analysis of forced oscillations; however, the standard SINDy approach is known to be sensitive to measurement noise. Furthermore, ML has been leveraged for the optimal tuning of Power System Stabilizers (PSS): deep reinforcement learning has been utilized to optimize PSS parameters~\cite{zhang2020deep,he2024measurement}, while hybrid deep learning models have been introduced to enhance overall control performance~\cite{sarkar2024fractional}.

\subsection{Paper contribution}

In this paper, we propose a novel deep learning framework for the identification of electromechanical oscillations in power systems, leveraging the SINDy-SENDAI architecture. This approach mitigates the inherent sensitivity of SINDy to measurement noise by integrating the SENDAI architecture to effectively decouple low- and high-frequency dynamics. The main contributions of this work are summarized as follows:\\

\begin{itemize}
    \item \textbf{Hierarchical Multiscale Framework:} This paper introduces the SINDy-SENDAI architecture, a novel framework that integrates SINDy-regularized latent dynamics with a hierarchical frequency extraction mechanism. This approach enables the high-fidelity decomposition of multi-bandwidth PMU data, effectively separating high-frequency transients from dominant system oscillations within a structured multiscale environment.

    \item \textbf{Parsimonious Latent Dynamics and Modal Extraction:} The proposed methodology discovers parsimonious governing equations within a learned latent space, facilitating the direct analytical extraction of critical electromechanical parameters, such as oscillation frequencies, damping ratios, and spatial mode shapes. This ensures both the interpretability of the identified model and a deep physical insight into the system stability.

    \item \textbf{Comprehensive Empirical Validation:} The robustness of SINDy-SENDAI is rigorously validated using the Kundur two-area benchmark and two large-scale real-world datasets: the 2016 Iberian power system oscillatory event and ambient measurements from the Italian power grid in 2021. Comparative results demonstrate that the proposed method outperforms state-of-the-art techniques in terms of reconstruction accuracy and predicted stability.\\
\end{itemize}


The remainder of this paper is organized as follows. Section~\ref{sec:SINDy_SENDAI} presents the SINDy-SENDAI architecture. Section~\ref{sec:results} reports the numerical results obtained from different simulated and real-world test cases. The limitations and future works are presented in Section \ref{sec:limitations} and, finally, Section~\ref{sec:conclusions} summarizes the main conclusions of this work.

\section{SINDy-SENDAI algorithm}\label{sec:SINDy_SENDAI}
\begin{figure*}[t]
    \centering
    \includegraphics[width=.97\linewidth]{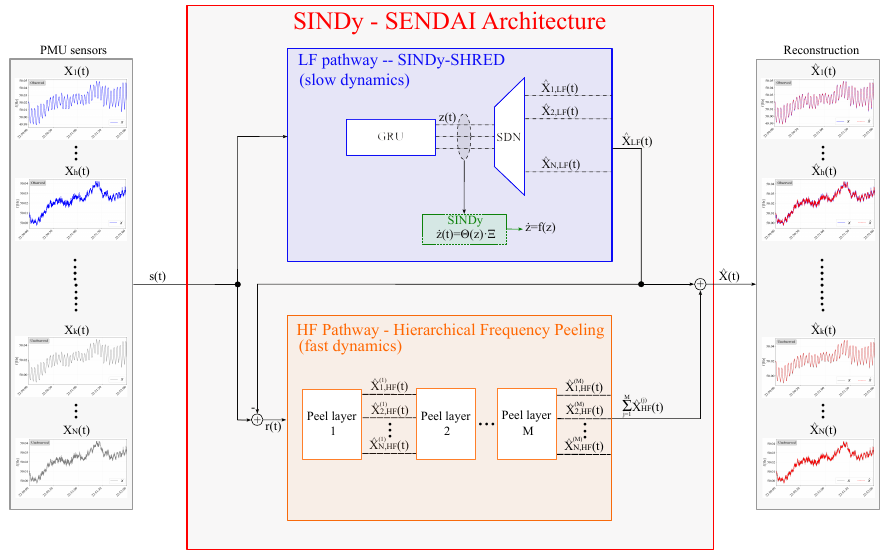}
    \caption{SINDy-SENDAI architecture. The framework consists of two pathways. The LF pathway takes the \textbf{s}(t) observed measurements (from $X_1(t)$ to $X_h(t)$) as input and outputs the reconstruction of the dominant low-frequency dynamics, denoted as $\hat{X}_{LF}(t)$, for all the sensors. By integrating the SINDy module, this pathway effectively captures the LF behavior and identifies a dynamical model $f(z)$ in the latent space. The HF pathway, instead, receives as input the residuals computed as the difference between the original sensor measurements \textbf{s}(t) and the LF reconstruction $\hat{X}_{LF}(t)$. By combining the LF and HF pathways, SINDy-SENDAI reconstructs the complete set of sensor signals, both observed and unobserved, yielding the final output $\hat{X}(t)$.}
    \label{fig:SINDy_SENDAI}
\end{figure*}

Power system dynamics inherently span multiple temporal scales, ranging from fast electromagnetic transients (microseconds to milliseconds) to electromechanical oscillations (seconds) and long-term operational dynamics (minutes to hours). These heterogeneous time scales arise from the interaction of electrical network components, rotating machines, control systems, and market-driven operational decisions \cite{hatziargyriou2020definition}. As a result, no single modeling framework can accurately capture all dynamic phenomena simultaneously and accurately. A multiscale modeling approach is therefore essential to ensure modeling efficiency while preserving physical accuracy \cite{gao2009frequency}.\\
Reconstructing full system dynamics and identifying governing equations from sparse PMU measurements necessitate an architecture capable of effectively isolating and capturing distinct spectral components. To address this, we propose the SINDy-SENDAI framework illustrated in Fig. \ref{fig:SINDy_SENDAI}.

SENDAI~\cite{zhang2026sendai} is specifically designed for the multiscale reconstruction problem. SENDAI decomposes the system state into a Low-Frequency (LF) pathway, which captures the dominant slow dynamics, and a hierarchical High-Frequency (HF) pathway composed of multiple sequential correction layers that progressively refine the reconstruction, as expressed by
\begin{equation}
\hat{\mathbf{x}}(t) = \hat{\mathbf{x}}_{\mathrm{LF}}(t)
  + \sum_{j=1}^{M}
    \hat{\mathbf{x}}_{\mathrm{HF}}^{(j)}(t),
\label{eq:sendai_decomp}
\end{equation}
where $\hat{\mathbf{x}}(t)$ denotes the reconstructed system state vector, $\hat{\mathbf{x}}_{\mathrm{LF}}(t)$ represents the reconstructed LF component, and $\hat{\mathbf{x}}_{\mathrm{HF}}^{(j)}(t)$ corresponds to the $j$-th correction (peel) layer of the HF pathway. Each HF peel layer operates on the residual of the sensor measurements left by the preceding reconstruction stages, defined as
\begin{equation}\label{eq:sensor_residuals}
\mathbf{r}^{(j)}(t) = \mathbf{s}(t) - \mathbf{H}\hat{\mathbf{x}}^{(j-1)}(t),
\end{equation}
where $\mathbf{r}^{(j)}(t)$ is the residual provided as input to the $j$-th HF layer, $\mathbf{s}(t)$ is the vector of observed sensor measurements, and $\mathbf{H}$ is the observation matrix that maps the reconstructed system state to the measurement space. Through this hierarchical structure, SENDAI effectively separates distinct spectral components into interpretable and progressively smaller correction terms, thereby avoiding mode interference between low- and high-frequency dynamics.  The method is an improvement on previous scale separation methods~\cite{kutz2015multi,li2019discovering} which lack the capability of producing a joint model of a system. The LF backbone of SENDAI builds upon the SHallow REcurrent Decoder
(SHRED) architecture~\cite{williams2024sensing, pomarico2026shred}, which is grounded in Takens' embedding theorem~\cite{takens2006detecting}, and consists of a temporal encoder and a shallow decoder. The temporal unit maps a lagged window of sparse sensor observations, $\{\mathbf{s}(t), \mathbf{s}(t-1), \dots, \mathbf{s}(t-L)\}$, where $L$ denotes the lag length, into a low-dimensional latent state $\mathbf{z}(t) \in \mathbb{R}^{d_z}$. The shallow decoder then reconstructs the spatial state from this latent representation.

However, this work proposes an extended version of the SENDAI framework to enhance its interpretability and suitability for power system dynamic analysis. While SENDAI is inherently designed to separate LF and HF dynamics, we augment the LF pathway by embedding a SINDy module~\cite{brunton2016discovering,gao2025sparse} directly within the latent space. This extension enables the identification of an explicit dynamical model governing the LF latent variables, thereby transforming the architecture from a purely data-driven reconstruction tool into a physics-informed and interpretable framework. To enforce dynamical structure in the LF latent space, SINDy is applied to identify a parsimonious ordinary differential equation of the form
\begin{equation}
\dot{\mathbf{z}} = \boldsymbol{\Xi}^\top \boldsymbol{\Theta}(\mathbf{z}),
\label{eq:sindy_ode}
\end{equation}
where $\boldsymbol{\Theta}(\mathbf{z})$ denotes a library of candidate nonlinear functions, and $\boldsymbol{\Xi}$ is a sparse coefficient matrix determining the active terms in the model. Within this formulation, the latent trajectories are jointly optimized to ensure accurate reconstruction of the observed data while simultaneously conforming to a sparse and interpretable dynamical system. As a result, the learned latent space is endowed with explicit governing equations, enabling modal analysis, dynamic stability assessment of LF dynamics, and forward time integration for extrapolation beyond the observation window. As a result, SINDy-SENDAI simultaneously reconstructs the system state and identifies parsimonious governing equations for the learned latent dynamics, thereby combining accurate data-driven estimation with interpretable dynamical modeling.

\subsection{Mathematical formulation}
This section presents the mathematical formulation of the proposed SINDy-SENDAI architecture, detailing both the latent-space encoding-decoding process and the integration of the SINDy module for interpretable dynamic modeling.

\subsubsection{\textbf{Low-Frequency Pathway}} \label{sec:lf_pathway}

The LF pathway leverages the SINDy-SHRED architecture introduced in~\cite{gao2025sparse}, which jointly trains a temporal unit (e.g. GRU~\cite{dey2017gate} or LSTM network~\cite{cho2014learning}) with a shallow decoder with SINDy regularization on latent space dynamics. \\
Let $\mathbf{x}(t) \in \mathbb{R}^N$ denote the all the $N$ PMUs measurements, and let $\mathbf{s}(t) = \mathbf{H}\mathbf{x}(t) \in \mathbb{R}^p$ represent a subset of $p$ PMUs, with $\mathbf{H} \in \mathbb{R}^{p \times N}$ being the observation matrix and $p \ll N$. \\
Given a lagged window of sensor observations 
\[
\mathbf{S}_{t-L:t} = [\mathbf{s}(t-L+1), \dots, \mathbf{s}(t)] \in \mathbb{R}^{L \times p},
\] 
the multi-layer GRU $\mathcal{E}$ maps this sequence into the latent space $\mathbf{z}$ as
\begin{equation}
\mathbf{z}(t) = \mathcal{E}(\mathbf{S}_{t-L:t}; \, \theta_{\mathrm{enc}}) \in \mathbb{R}^{d_z},
\label{eq:encoder}
\end{equation}
where $\theta_{\mathrm{enc}}$ denotes the trainable parameters and $d_z \ll N$ is the dimension of the latent space.\\
Then, a shallow Multi-Layer Perceptron (MLP) decoder $\mathcal{D}$ maps the latent space back to the full state space:
\begin{equation}
\hat{\mathbf{x}}_{\mathrm{LF}}(t) = \mathcal{D}(\mathbf{z}(t);\, \theta_{\mathrm{dec}}) \in \mathbb{R}^n.
\label{eq:decoder}
\end{equation}
here, $\theta_{\mathrm{dec}}$ denotes the trainable shallow decoder parameters.

Furthermore, the latent trajectory $\mathbf{z}(t)$ is constrained to follow the sparse dynamical model in~\eqref{eq:sindy_ode} through a dedicated regularization loss during training. Instead of matching derivatives directly, the SINDy loss is computed via sub-stepped Euler integration. Given consecutive latent states $\mathbf{z}(t)$ and $\mathbf{z}(t+\Delta t)$, the predicted next state is
\begin{equation}
\hat{\mathbf{z}}(t+\Delta t) = \mathbf{z}(t) + \sum_{j=1}^{N_s} \frac{\Delta t}{N_s}\, \boldsymbol{\Xi}^\top \boldsymbol{\Theta}\!\left(\hat{\mathbf{z}}^{(k)}\right),
\label{eq:euler_sindy}
\end{equation}
where $N_s$ denotes the number of sub-steps and $\hat{\mathbf{z}}^{(k)}$ is the intermediate latent state after $j$ sub-steps. The SINDy loss is defined as:
\begin{equation}
\mathcal{L}_{\mathrm{SINDy}} = \left\lVert \hat{\mathbf{z}}(t+\Delta t) - \mathbf{z}(t+\Delta t) \right\rVert_2^2,
\end{equation}
quantifying discrepancies between predicted and inferred latent states. As demonstrated in \cite{gao2025sparse}, this integration-based formulation captures dynamics near bifurcations and transitions more effectively than direct derivative matching. The total objective for the LF pathway is:
\begin{equation}
\mathcal{L}_{\mathrm{LF}} = \mathcal{L}_{\mathrm{recon}} + \lambda_{\mathrm{SINDy}}\, \mathcal{L}_{\mathrm{SINDy}},
\label{eq:lf_loss}
\end{equation}
where $\mathcal{L}_{\mathrm{recon}}$ is the reconstruction loss and $\lambda_{\mathrm{SINDy}}$ is increased via a warmup schedule to first establish a stable latent representation. Sparsity in $\boldsymbol{\Xi}$ is enforced through iterative thresholding \cite{champion2019data,mars2024bayesian}, which prunes minor coefficients and refits active terms. Finally, once the LF pathway has converged, a post-hoc SINDy identification is performed: the trained encoder is used to extract the latent trajectory $\mathbf{z}(t)$ over the entire training window, and a fresh sparse regression is fit to these latents via sequentially thresholded least squares, yielding a refined coefficient matrix $\boldsymbol{\Xi}$. In contrast to the SINDy regularization applied during training, this step decouples model discovery from representation learning, producing the final governing equations~\eqref{eq:sindy_ode} employed for forecasting and stability analysis. The identified model is verified for stability by ensuring that its autonomous integration produces bounded trajectories.

\subsubsection{\textbf{Hierarchical High-Frequency Peeling}}
\label{sec:hf_peeling}

Once the LF pathway has been trained, it provides a smooth baseline reconstruction that captures the dominant slow dynamics. The remaining discrepancy between the LF prediction and the actual sensor measurements contains higher-frequency content associated with fast dynamic phenomena. Following the hierarchical peeling strategy of SENDAI~\cite{zhang2026sendai}, a sequence of $M$ correction layers is applied, as described in~\eqref{eq:sendai_decomp}. Denoting $\hat{\mathbf{x}}^{(0)} = \hat{\mathbf{x}}_{\mathrm{LF}}$, each layer $j$ performs the following steps:
\begin{itemize}
    \item[(i)] Computes the sensor residuals as \eqref{eq:sensor_residuals};
    
    \item[(ii)] Applies the $j$-th high-frequency correction network:
    \begin{equation} \label{eq:hf_correction}
        \Delta\hat{\mathbf{x}}_{\mathrm{HF}}^{(j)} = \gamma^{(j)}\, \mathcal{H}^{(j)}(\mathbf{r}^{(j)}; \theta_{\mathrm{HF}}^{(j)}),
    \end{equation}
    where $\mathcal{H}^{(j)}$ is the correction network parameterized by $\theta_{\mathrm{HF}}^{(j)}$.
    \item[(iii)] Updates the reconstruction:
    \begin{equation}
        \hat{\mathbf{x}}^{(j)} = \hat{\mathbf{x}}^{(j-1)} + \Delta\hat{\mathbf{x}}_{\mathrm{HF}}^{(j)}.
    \end{equation}
\end{itemize}
Each correction layer is trained sequentially, with all preceding layers frozen, ensuring that it exclusively models the residual dynamics not captured by prior corrections.


Each correction network $\mathcal{H}^{(j)}$ operates on a short temporal window of $T_{\mathrm{HF}}$ consecutive sensor residuals $\mathbf{R}^{(j)} \in \mathbb{R}^{T_{\mathrm{HF}} \times p}$. A one-dimensional convolutional encoder extracts temporal features from $\mathbf{R}^{(j)}$, which are then passed to a per-timestep MLP decoder that predicts full-state corrections for each time step in the window. As shown in~\eqref{eq:hf_correction}, the output of each correction network is modulated by a learnable scalar $\gamma^{(j)}$. This parameter is initialized to a small value that decreases with the layer index $j$, regulating the magnitude of each layer's correction and promoting stable training.

A distinguishing feature of the present framework is the enforcement of sparsity in the \emph{temporal} frequency domain. Unlike the original SENDAI formulation which regularizes corrections in the spatial frequency domain for two-dimensional fields, SINDy-SENDAI operates on one-dimensional PMU time series where the relevant structure is temporal. For each HF layer, the discrete Fourier transform of the correction signal is computed along the temporal axis, yielding a set of frequency components that are partitioned into an in-band region (frequencies up to a prescribed physical limit $f_{\max}$, chosen to encompass the target oscillation range) and an out-of-band region. The sparsity regularizer is
\begin{equation}
\mathcal{R}_{\mathrm{sparse}}^{(j)} = \frac{\lVert
\hat{\mathbf{c}}_{\mathrm{in}}^{(j)} \rVert_1}{\lVert
\hat{\mathbf{c}}_{\mathrm{in}}^{(j)} \rVert_2 + \epsilon} + \beta\,
\frac{\lVert \hat{\mathbf{c}}_{\mathrm{out}}^{(j)}
\rVert_2^2}{\lVert \hat{\mathbf{c}}^{(j)} \rVert_2^2 +
\epsilon}\,,
\label{eq:temporal_sparsity}
\end{equation}
where $\hat{\mathbf{c}}_{\mathrm{in}}^{(j)}$ and $\hat{\mathbf{c}}_{\mathrm{out}}^{(j)}$ denote the magnitudes of the in-band and out-of-band temporal Fourier coefficients, respectively, and $\beta \gg 1$ strongly penalizes out-of-band leakage. The first term is the ratio between the $\ell_1$ and $\ell_2$ norms of the in-band Fourier magnitudes, a scale-invariant sparsity surrogate. Denoting by $n$ the number of in-band frequency components, this ratio is bounded in $[1, \sqrt{n}]$: it attains its minimum when a single component is active and its maximum when energy is spread uniformly across all in-band components. Minimizing it thus concentrates the spectral energy of each correction into a few dominant frequencies, encouraging each peeling layer to capture a small number of coherent oscillatory modes.

The training objective for each HF peeling layer $j$ balances reconstruction accuracy, sparsity, and magnitude control is:
\begin{equation}
\mathcal{L}_{\mathrm{HF}}^{(j)} = \mathcal{L}_{\mathrm{recon}}^{(j)} + \lambda_{\mathrm{sp}}\, \mathcal{R}_{\mathrm{sparse}}^{(j)} + \lambda_{\mathrm{mag}}\, \mathcal{L}_{\mathrm{mag}}^{(j)},
\label{eq:hf_loss}
\end{equation}
where $\mathcal{L}_{\mathrm{recon}}^{(j)}$ minimizes the mismatch between predicted corrections and observed residuals, while $\mathcal{L}_{\mathrm{mag}}^{(j)}$ penalizes excessive corrections to prevent layer dominance. The peeling layers are trained sequentially; upon completing layer $j$, sensor residuals are recomputed to incorporate cumulative corrections from layers $1$ through $j$, providing the target for the subsequent layer. To optimize the trade-off between reconstruction fidelity and spectral interpretability, the sparsity weight $\lambda_{\mathrm{sp}}$ is gradually increased during a warmup phase, followed by a fine-tuning stage with a reduced penalty.

\subsubsection{\textbf{Test-time inference and Autonomous Forecasting}}
\label{sec:inference}

At test-time inference, the reconstruction proceeds in two stages. Given a window of PMU measurements $\mathbf{S}_{t-L:t}$ and current sensor readings $\mathbf{s}(t)$, the LF temporal unit maps the sensor history to the latent space $\mathbf{z}(t)$ via~\eqref{eq:encoder}, and the LF decoder produces the baseline reconstruction $\hat{\mathbf{x}}_{\mathrm{LF}}(t)$ via~\eqref{eq:decoder}. The HF pathway then iteratively refines this estimate: for each peeling layer $\ell = 1, \ldots, N$, the sensor residual is computed and the correction applied, yielding the final reconstruction as in~\eqref{eq:sendai_decomp}.



A distinctive capability enabled by the SINDy-regularized latent space is the autonomous forward prediction beyond the sensor observation window. Once the post-hoc SINDy model has been identified, the ODE defined as in~\eqref{eq:sindy_ode} can be integrated forward from the last observed latent state, and the decoder applied to produce full state predictions without further sensor input:
\begin{equation}
\hat{\mathbf{x}}_{\mathrm{pred}}(t+\tau) = \mathcal{D}\!\left(\mathbf{z}_{\mathrm{SINDy}}(t+\tau)\right), \quad \tau > 0,
\label{eq:sindy_forecast}
\end{equation}
where $\mathbf{z}_{\mathrm{SINDy}}(t+\tau)$ is obtained by numerical integration of~\eqref{eq:sindy_ode} from the initial condition $\mathbf{z}(t)$. This procedure yields a LF forecast that captures the slow dynamical trends of the system without requiring HF corrections, as the latter depend on sensor residuals unavailable beyond the observation window. The spectral separation enforced during training ensures that the identified ODE represents a parsimonious dynamical system~\cite{kutz2022parsimony}, rather than the full multiscale complexity of the power grid, which promotes stable and accurate autonomous integration.  Dynamical coupling across scales is not included even though such couplings, especially nonlinear couplings can produce non-trivial electromagnetic interactions~\cite{proctor2005passive}.  This will be considered in future work.

\subsubsection{\textbf{Stability Assessment and Mode shapes}} 
Once the SINDy model is identified, classical stability analysis can be performed in the latent space. Assuming a linear state-space representation characterized by the system matrix $\mathbf{A} \in \mathbb{R}^{d_z \times d_z}$, the dynamic stability is assessed via the eigendecomposition of $\mathbf{A}$. Since the latent representation captures the dominant system dynamics, $\mathbf{A}$
inherently contains the electromechanical modes of the power system. Consequently, the eigenvalues $\lambda_i = \sigma_i \pm j\omega_i$ provide direct information on modal frequency $f_i = \omega_i / (2\pi)$ and damping ratio $\xi_i = -\sigma_i / \sqrt{\sigma_i^2 + \omega_i^2}$. Furthermore, the eigenvectors $\boldsymbol{\mu}_i$ can be used to reconstruct the mode shapes (i.e., spatial patterns across PMU space). For each conjugate eigenvalue pair, a latent trajectory isolating that mode is synthesized as $\mathbf{z}_i(t) = \alpha_i\,\mathrm{Re}(e^{\lambda_i t}\,\boldsymbol{\mu}_i)$, where $\alpha_i$ is a scaling factor, and decoded through the frozen LF decoder over multiple time steps, yielding the time evolution of that mode in the PMU measurement space. The amplitude and phase of the decoded waveform at each PMU location are then extracted via spectral analysis, providing the complex mode shape associated with each identified oscillatory mode.

\section{Numerical results}\label{sec:results}
This section presents a comprehensive evaluation of the proposed method through three case studies, encompassing both simulated scenarios and real-world European oscillatory events. The SINDy-SENDAI architecture is kept identical across all case studies and is configured as follows\footnote{The hyperparameter tuning was performed based on the authors' experience and engineering judgment. Future work will investigate automated hyperparameter optimization techniques to improve model performance, robustness, and applicability in real-time settings.}:
\begin{itemize}
    \item \textbf{LF pathway:} The temporal encoder is a GRU with two hidden layers of 64 neurons each and a lag length of 40~seconds. The shallow decoder is implemented as a MLP with two hidden layers of 256 neurons. The latent space is four-dimensional, and the SINDy model identified in the latent space is constrained to be linear. The LF pathway is designed to identify dynamics within the electromechanical oscillation bandwidth (0.1--3~Hz).
    
    \item \textbf{HF pathway:} The HF correction module consists of a one-dimensional convolutional neural network (1D-CNN)\cite{li2021survey} with two layers comprising 32 and 64 filters, respectively, and a kernel size of 5. The convolutional encoder is followed by an MLP decoder with two hidden layers of 128 neurons each. The number of peeling layers is set to 10.
\end{itemize}
All the datasets are partitioned into training, validation, and test subsets using a 70\%–15\%–15\% split, respectively. However, detailed information regarding the code used in this work are openly available through the GitHub repository referenced at the end of the paper.

\subsection{Benchmark} \label{sec:benchmarks}
\begin{figure}[htbp]
    \centering
    \includegraphics[width=0.9\linewidth]{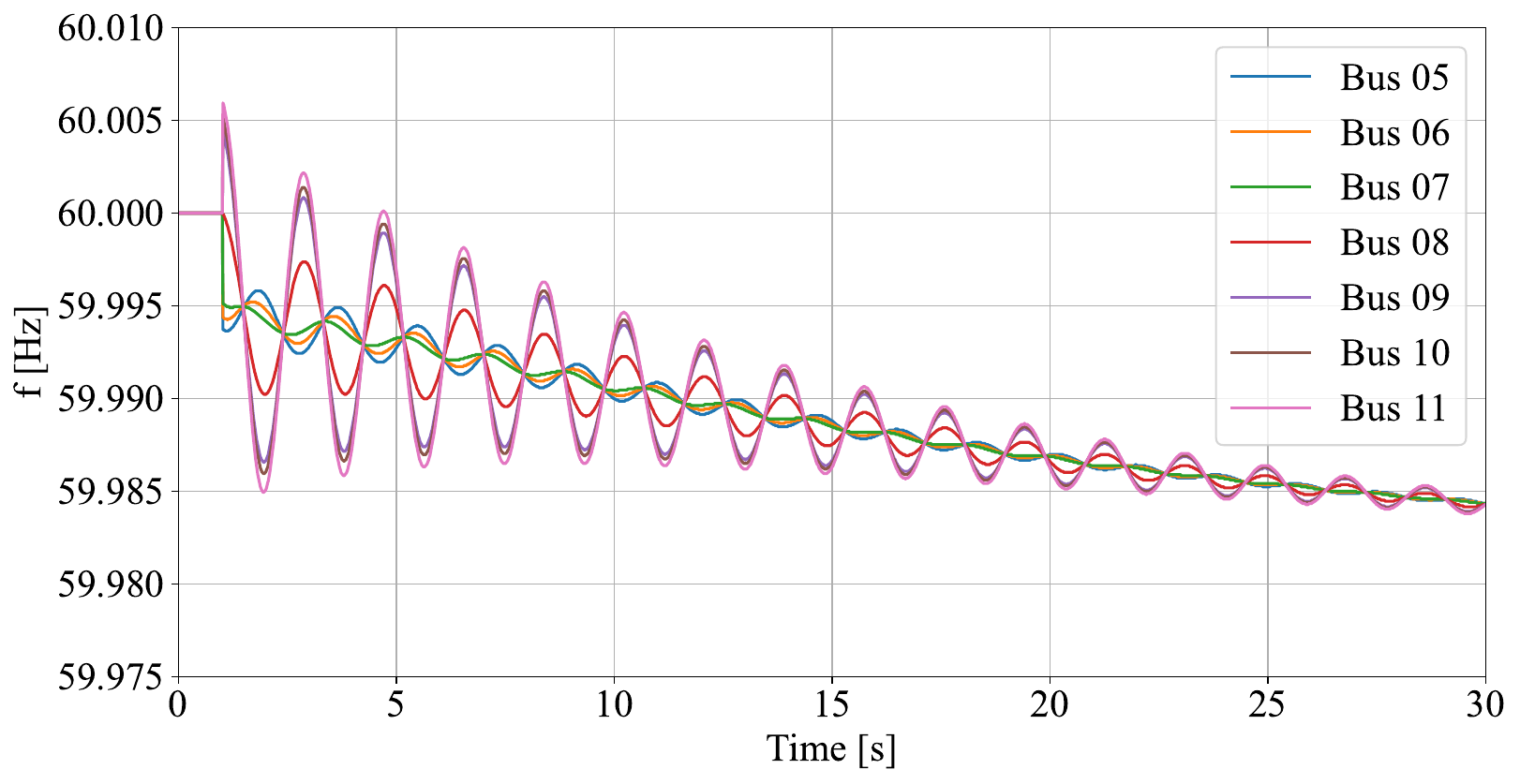}
    \caption{Frequency measurements recorded during the interarea oscillation simulation in the two-area Kundur system.}
    \label{fig:Two_Area_kundur_frequency}
\end{figure}

This section presents a simulation-based analysis to evaluate the performance of the proposed SINDy-SENDAI approach in comparison with the state-of-the-art hDMD method~\cite{vicario2022practical}, using conventional Modal Analysis (MA) as the reference benchmark. The well-known two-area Kundur system~\cite{kundur2007power} is adopted as the test system. It exhibits a characteristic interarea oscillation at \SI{0.5431}{\hertz} with a damping ratio of \SI{2.851}{\percent}. Since generator rotor speeds are not directly available in real-world WAMS applications, simulated frequency measurements are used to emulate a realistic monitoring scenario. The oscillation is triggered by decreasing the mechanical torque in Area~1 and increasing it in Area~2. Frequency measurements are sampled every \SI{20}{\milli\second}, as shown in Fig.~\ref{fig:Two_Area_kundur_frequency}.
The PMU measurements serve as input to the proposed SINDy-SENDAI approach, and for dynamic stability analysis, the linear SINDy model identified in the latent space is employed. Specifically, the four ODEs discovered by SINDy are:
\begin{equation}
\resizebox{0.89\columnwidth}{!}{$
\begin{cases}
\dot{z}_1 = - 27.479\, z_1 + 30.160\, z_2 - 36.745\, z_3 + 10.723\, z_4 + 5.262,\\[2mm]
\dot{z}_2 = - 9.567\, z_1 + 11.806\, z_2 - 14.445\, z_3 + 5.317\, z_4 + 1.624,\\[1mm]
\dot{z}_3 = 5.680\, z_1 - 4.299\, z_2 + 5.382\, z_3 - 0.856\, z_4 - 1.221,\\[1mm]
\dot{z}_4 = - 24.335\, z_1 + 26.797\, z_2 - 32.789\, z_3 + 10.192\, z_4 + 4.541
\end{cases}
$}
\end{equation}

Since the discovered model is linear, an eigendecomposition analysis can be performed to extract the oscillatory modes and compute the corresponding modal frequency $f$ and damping ratio $\xi$. The relative percentage error of the modal parameters estimated by the data-driven approaches with respect to conventional MA is computed as
\begin{equation}
\varepsilon_{r}^{\mathcal{X}}[\%] = 
\frac{\mathcal{X}^{\mathcal{Y}} - \mathcal{X}^{MA}}{\mathcal{X}^{MA}} \cdot 100,
\label{eq:relative_error}
\end{equation}
where $\mathcal{X} = \{f,\xi\}$ and $\mathcal{Y} = \{\text{hDMD}, \text{SINDy-SENDAI}\}$ denote the method used. The results of this comparison are reported in Table~\ref{tab:kundur}.
\begin{table}[htbp]
\caption{Modal parameters results comparison for the two-area Kundur system.}
\label{tab:kundur}
\centering
\begin{tabular}{c|cc|cc|cc}
\toprule
\toprule
 & \multicolumn{2}{c|}{\textbf{MA}} 
 & \multicolumn{2}{c|}{\textbf{hDMD}} 
 & \multicolumn{2}{c}{\textbf{SINDy-SENDAI}} \\
 \midrule
$f$ [Hz] / $\xi$ [\%] 
& 0.5431 & 2.851 
& 0.52   & 3.915
& 0.549  & 2.38 \\
$\varepsilon_{r}^{f}$ [\%] / $\varepsilon_{r}^{\xi}$ [\%] 
& - & - 
& -4.25 & 37.32 
& 1.08  & -0.19 \\
\bottomrule
\bottomrule
\end{tabular}
\end{table}
As can be observed, the proposed SINDy-SENDAI approach outperforms hDMD in both modal frequency and damping ratio estimation.

\subsection{Oscillatory Event on December 1\textsuperscript{st}, 2016} \label{sec:ENSTOE2016}
\begin{figure}[htbp]
    \centering
    \includegraphics[width=0.95\linewidth]{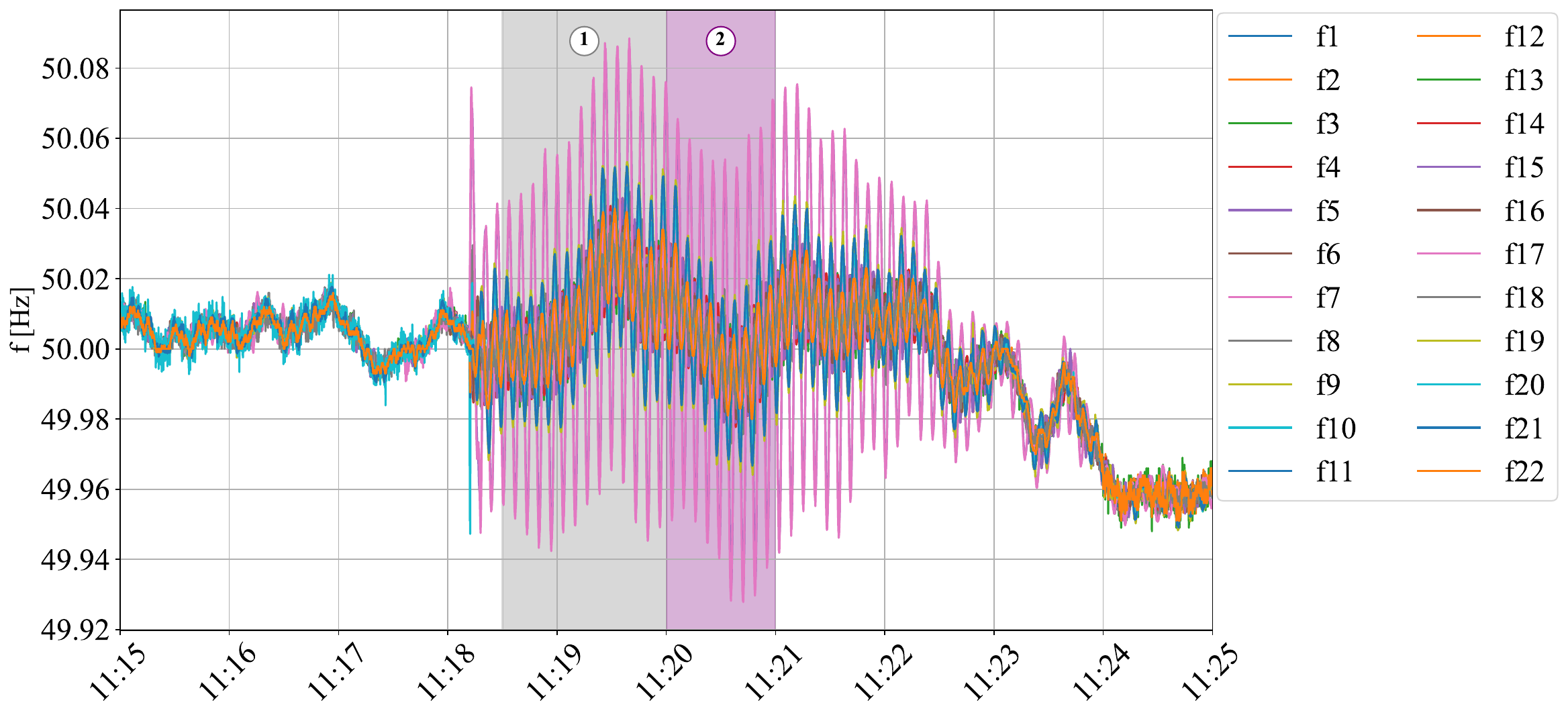}
    \caption{PMU measurements of the oscillatory event on December~1\textsuperscript{st}, 2016.}
    \label{fig:ENSTOE_2016}
\end{figure}

This section presents the application of the proposed approach to a real oscillatory event that occurred in the European power system on December~1\textsuperscript{st}, 2016. The event consisted of a significant interarea oscillation involving the Iberian Peninsula, the Balkan region, and Central Europe, commonly referred to as the East–Central–West mode~\cite{entsoe2017}. The frequency measurements recorded across the European power system, with a sampling frequency $f_s$ of 10 Hz, are shown in Fig.~\ref{fig:ENSTOE_2016}. The disturbance was characterized by a dominant interarea oscillatory mode at approximately \SI{0.15}{\hertz}.

\begin{figure*}[htbp]
    \centering
    \includegraphics[width=1\linewidth]{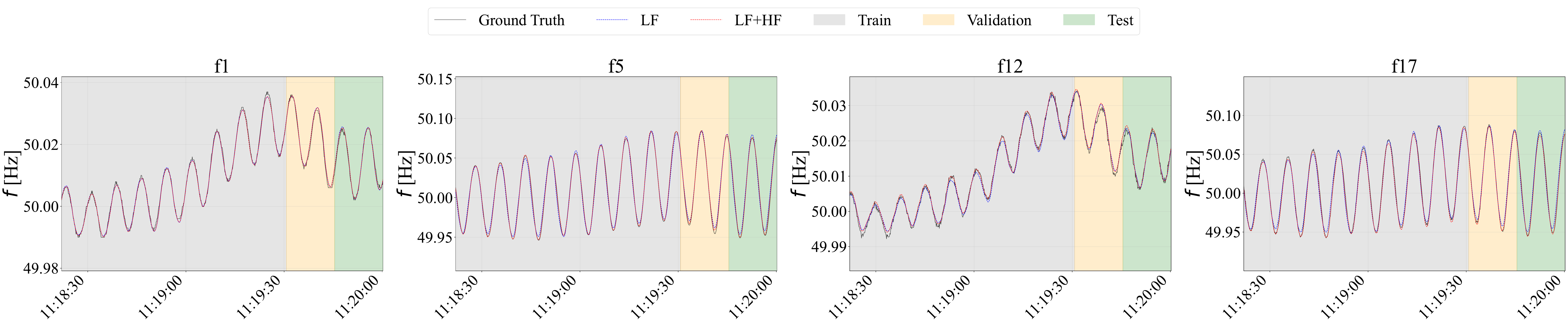}
    \caption{Reconstruction of the LF pathway and the combined LF+HF pathways across $f_1$, $f_5$, $f_{12}$ and $f_{17}$ during the time window from 11{:}18{:}30 to 11{:}20{:}00 using the SINDy-SENDAI architecture. When an interarea oscillation is excited, the LF pathway alone already provides an accurate reconstruction of the dominant dynamics, while the addition of the HF pathway further refines the full estimation.}
    \label{fig:PMU2016_recon_w1}
\end{figure*}

\begin{figure}[htbp]
    \centering
    \includegraphics[width=0.9\linewidth]{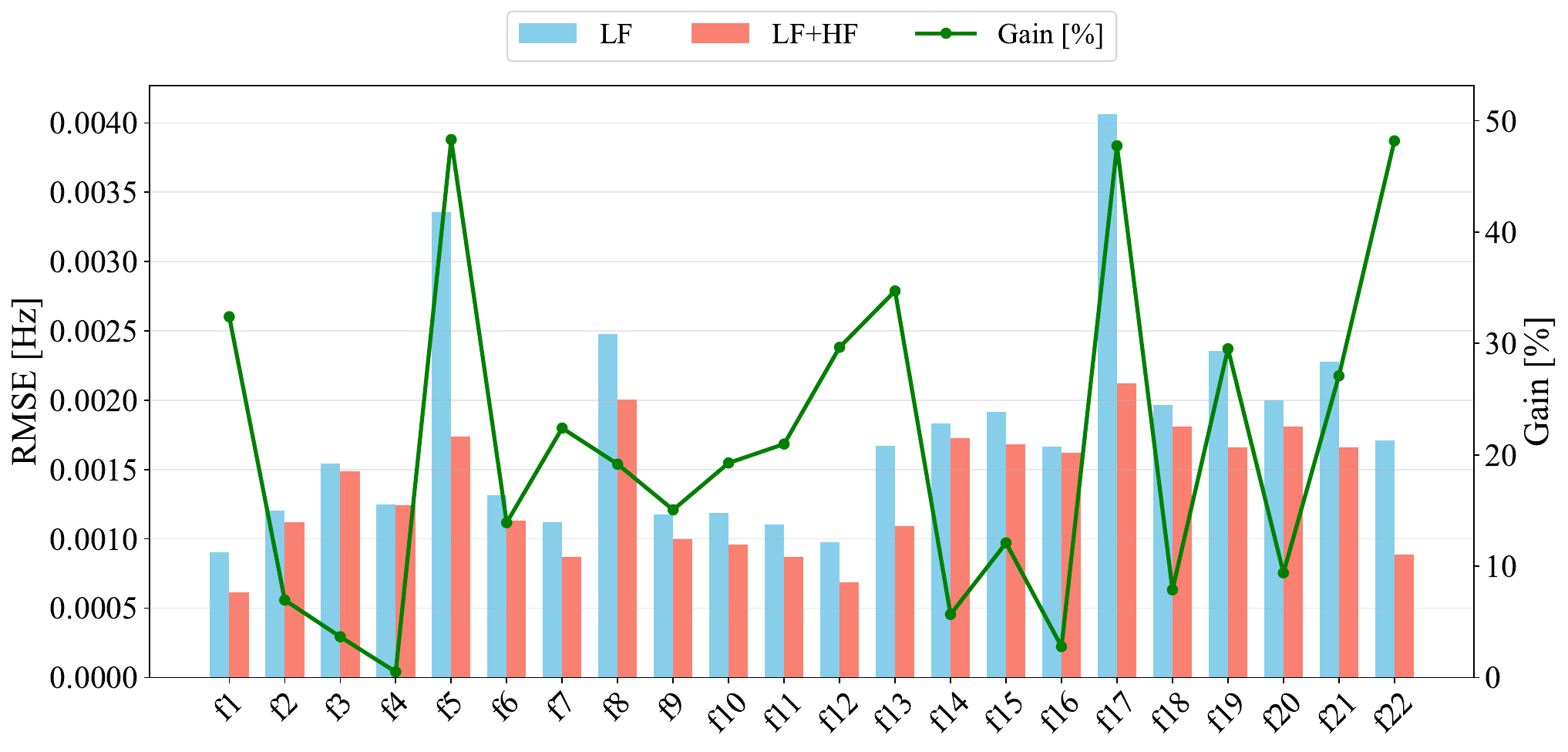}
    \caption{RMSE [Hz] and Gain [\%] between LF and LF+HF for all the PMUs reconstructions of the oscillatory event in 2016.}
    \label{fig:RMSE_Gain_2016}
\end{figure}

First, we illustrate how the proposed approach would operate in the time window highlighted in Fig.~\ref{fig:ENSTOE_2016}. Specifically, we assume to be at time $t = 11{:}20{:}00$ during the oscillatory event and use the preceding 90 seconds of data to train, validate, and test the SINDy-SENDAI model. Once the SINDy model governing the latent variables $\mathbf{z}(t)$ is identified, the discovered ODE in~\eqref{eq:sindy_ode} is autonomously integrated forward to forecast the interarea mode one minute ahead.\\
For a sake of compactness, figure~\ref{fig:PMU2016_recon_w1} illustrates the reconstruction performance for the PMUs exhibiting the lowest and highest reconstruction errors. The results, shown over the training, validation, and test intervals, demonstrate the ability of SINDy-SENDAI to accurately reproduce the measured dynamics across the entire observation period, even for the most challenging PMU signals.
The LF pathway alone captures the dominant oscillatory dynamics with high fidelity, while the inclusion of the HF pathway further improves the overall reconstruction by refining the remaining HF components.
To further quantify the reconstruction performance, the Root Mean Squared Error (RMSE) is calculated as
\begin{equation}
    \resizebox{0.89\columnwidth}{!}{$
    f^{(i)}_{\mathrm{RMSE},m} = \sqrt{\frac{1}{n}\sum_{j=1}^{n}\left(f^{(i,j)}_{(\mathrm{SINDy\text{-}SENDAI},m)} - f^{(i,j)}_{\mathrm{ground\;truth}}\right)^2}
    $}
\end{equation}
where $i$ is the PMU index, $j$ indexes time samples, and $n$ denotes their total number; $m$ denotes either the LF or LF+HF pathway. Furthermore, the improvement gained by including the HF pathway is quantified by the percentage gain, defined as
\begin{equation}
    \mathrm{Gain} = \frac{f_\text{RMSE,LF}^{(i)} - f_\text{RMSE,LF+HF}^{(i)}}{f_\text{RMSE,LF}^{(i)}} \times 100,
\end{equation}
This metric quantifies the relative improvement in reconstruction accuracy obtained by incorporating the HF pathway alongside the LF pathway. Figure~\ref{fig:RMSE_Gain_2016} reports the RMSE values computed for all PMUs using both the LF and the combined LF+HF pathways, together with the corresponding gain achieved by incorporating the HF corrections.
As observed, for PMUs $f5$, $f17$, and $f22$, including the HF pathway in the reconstruction improves performance by nearly 50\%. In contrast, for the other PMUs, the LF pathway alone provides sufficiently accurate reconstruction, making the HF correction unnecessary.

\begin{figure*}[htbp]
    \centering
    \includegraphics[width=1\linewidth]{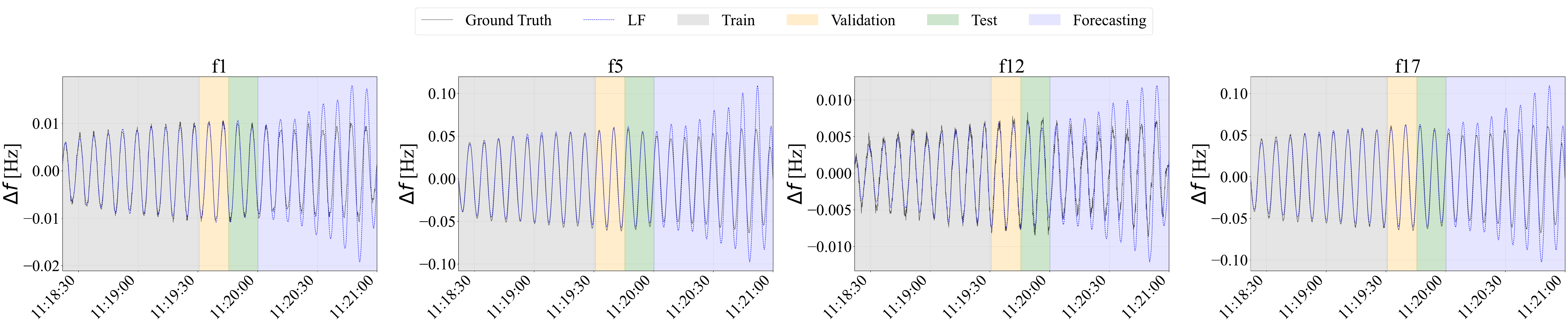}
    \caption{One-minute forecast of the LF pathway across $f_1$, $f_5$, $f_{12}$ and $f_{17}$. The SINDy model accurately reproduces the system dynamics during the first 30 seconds, after which errors gradually accumulate, leading to divergence from the true trajectory.}   
    \label{fig:PMU2016_inference_w1}
\end{figure*}

All the results presented so far focus on reconstructing the dynamics of the European power system across the train, validation and test dataset. However, as discussed in Section~\ref{sec:SINDy_SENDAI}, the SINDy-SENDAI architecture includes a SINDy module within the latent space to identify the governing equations underlying the LF mode. 
The linear system identified in the train time window by the SINDy module in the latent variables is:
\begin{equation}\label{eq:sindy_SPAIN}
\resizebox{0.89\columnwidth}{!}{$
\begin{cases}
    \dot{z_1}=-0.133 z_1 +1.556 z_2 -1.947 z_3 -0.184z_4+0.004\\
    \dot{z_2}=-0.334 z_1 +1.461 z_2 -1.835 z_3 -0.598z_4 +0.06\\
    \dot{z_3}=-0.240z_1 +2.724 z_2 -3.176z_3 -1.338z_4 +0.022\\
    \dot{z_4}=0.025z_1 -4.085z_2 +4.630z_3 +1.906z_4 +0.048
\end{cases}
$}
\end{equation}
Once the linear system is identified using the training dataset, it can be integrated with standard ODE solvers to forecast the near-future evolution of the LF pathway. Figure~\ref{fig:PMU2016_inference_w1} shows the forecasting results over a one-minute horizon, up to 11:21:00. It is important to note that the model does not capture the very low-frequency (mean) component of the system frequency dynamics, as this component is influenced by numerous external factors (e.g., primary and secondary frequency regulation).
During the initial phase, the forecast closely follows the ground truth, indicating that the identified SINDy model captures the dominant LF dynamics accurately. However, as the prediction horizon extends, small discrepancies accumulate because the model is not updated and remains fixed to the parameters identified during the training window. Consequently, the forecast trajectory gradually diverges from the true system state.  
To correct for this drift, the model must be updated, which can be achieved in two ways: (i) retraining the SINDy-SENDAI framework using new measurements, or (ii) re-initializing the integration of the SINDy ODE in~\eqref{eq:sindy_ode} from the current system state as a new starting point. The second approach is only valid if the underlying system dynamics have not changed significantly and the previously identified model remains a good approximation. If the dynamics have evolved substantially, retraining with updated data is necessary to ensure accurate forecasting.

Moreover, the dynamical model identified via SINDy in \eqref{eq:sindy_SPAIN} can be leveraged not only for forecasting purposes but also for dynamic stability analysis. By leveraging the linear structure of the identified latent system matrix, classical linear system theory is directly applicable. Specifically, the dominant electromechanical modes are extracted via eigendecomposition and compared against the hDMD applied in the full measurements space. The resulting frequency and damping ratio estimates for the training period are summarized in Table~\ref{tab:Iberian2016}.

\begin{table}[htbp]
\centering
\caption{Modal parameters estimation results in December 2016.}
\label{tab:Iberian2016}
\begin{tabular}{c|c|c}
\toprule
&\textbf{hDMD} & \textbf{SINDy-SENDAI} \\
\midrule
$f_1$ [Hz] & 0.151 & 0.150 \\
$\xi_1$ [\%] & -0.528 & -1.920 \\
\bottomrule
\end{tabular}
\end{table}

\begin{figure}[htbp]
    \centering
    \includegraphics[width=0.55\linewidth]{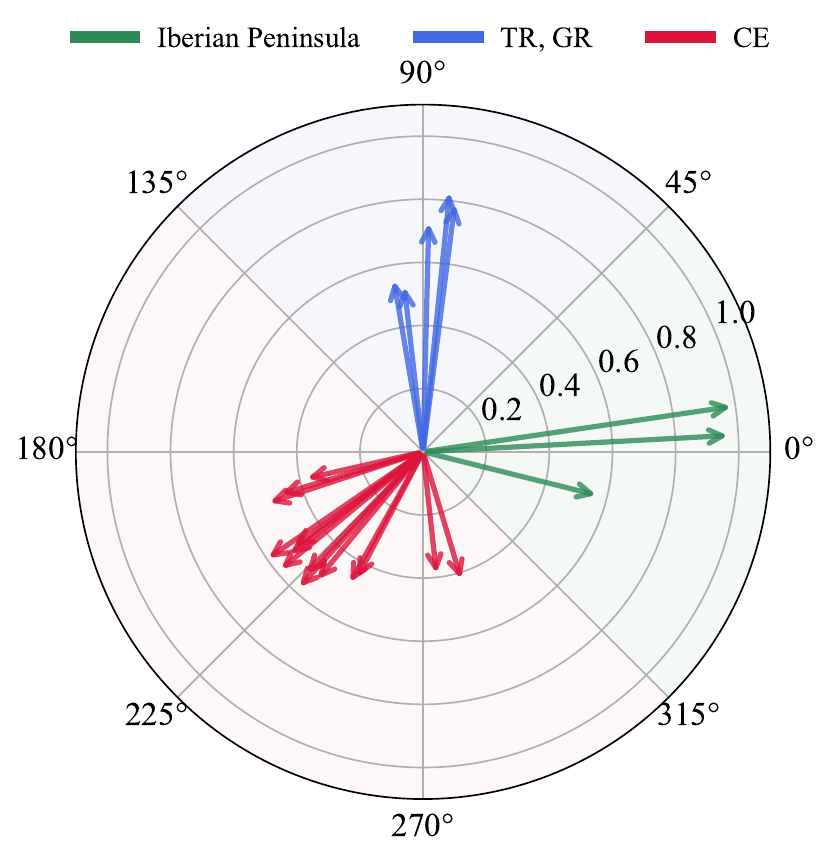}
    \caption{Mode shapes for the oscillatory event in December 2016.}
    \label{fig:mode_shapes_2016}
\end{figure}

The dominant oscillatory mode identified by both approaches exhibit frequency close to 0.15~Hz, corresponding to the well-known European interarea mode. Both methods yield highly consistent estimates of the modal frequencies, while slight differences are observed in the damping ratios. The true damping of the interarea mode cannot be directly determined, as this would require a complete modal analysis of the entire European power system, which is not feasible in practice. Lastly, the mode shapes, derived from the eigenvectors of the state matrix associated with the model in \eqref{eq:sindy_SPAIN} and mapped via the LF decoder (as detailed in Section~\ref{sec:SINDy_SENDAI}), are illustrated in Fig.~\ref{fig:mode_shapes_2016}.

As can be inferred, the proposed approach clearly identifies three coherent oscillatory regions: the Iberian Peninsula, the Greece–Turkey (GR-TR) area, and the Central European (CE) system. This clustering is consistent with the expected behavior of the European power system, confirming the effectiveness of the proposed method.
\subsection{Normal Grid Operation in March 2021}\label{sec:Marzo2021}
\begin{figure}[htbp]
    \centering
    \includegraphics[width=0.95\linewidth]{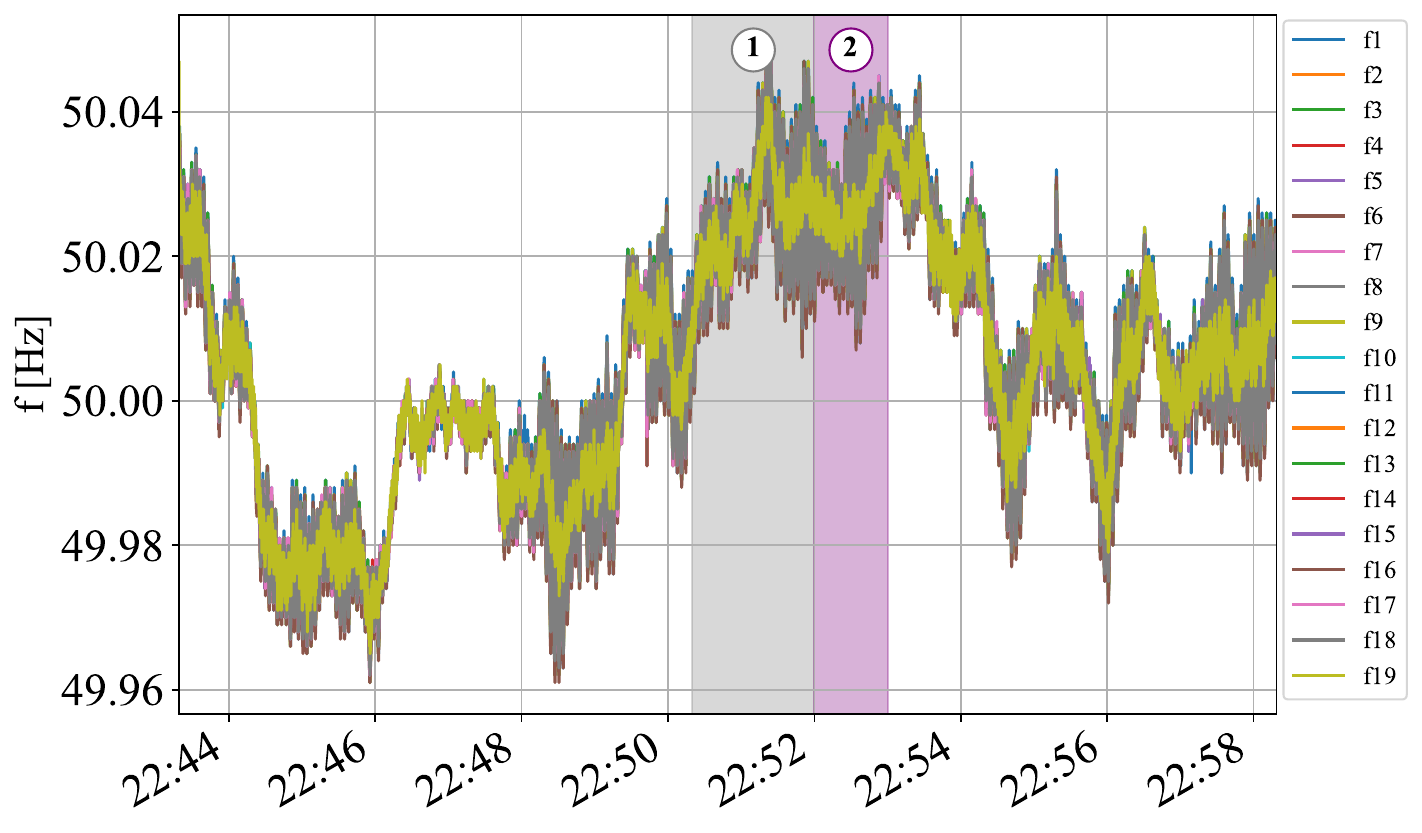}
    \caption{PMU measurements of the normal grid operation in March 2021.}
    \label{fig:Marzo_2021}
\end{figure}

\begin{figure*}[htbp]
    \centering
    \includegraphics[width=0.92\linewidth]{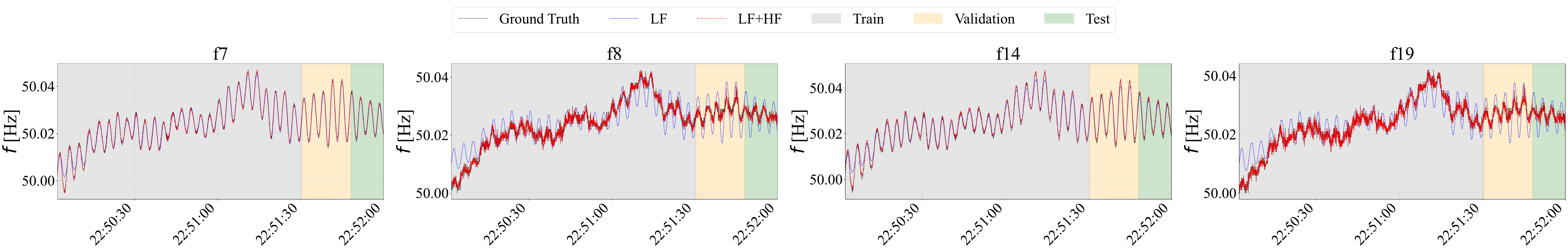}
    \caption{Visualization of the ground truth, the LF pathway, and the LF+HF pathways obtained with the SINDy-SENDAI approach for the measurement recordings ($f_7$, $f_8$, $f_{14}$ and $f_{19}$) within the time window highlighted in Fig.~\ref{fig:Marzo_2021}. As demonstrated, the LF pathway effectively identifies the interarea oscillatory component in each PMU signal, while the combination of LF and HF pathways enable accurate reconstruction of the ground-truth measurements.}
    \label{fig:PMU2021_recon_w1}
\end{figure*}

This section presents the results of the SINDy-SENDAI approach applied to real-world ambient data, where the interarea mode is known to be present but weakly excited, making it challenging to accurately identify and compute the corresponding modal parameters. To evaluate the method, we consider real-world PMU measurements (Fig.~\ref{fig:Marzo_2021}) located in the southern part of the Italian power system, sampled at a frequency of $f_s = 50$~Hz.
SINDy-SENDAI was applied to the time windows highlighted in Fig.~\ref{fig:Marzo_2021}. Specifically, the first time window \circled{1} was used for training, validation, and testing of the SINDy-SENDAI model, while the subsequent one-minute time window \circled{2} was reserved for forecasting the system dynamics. The reconstruction capability of SINDy-SHRED is clearly illustrated in Fig.~\ref{fig:PMU2021_recon_w1}, where the ground-truth PMU measurements are compared against the LF-only reconstruction and the combined LF+HF reconstruction. For the sake of compactness, only the PMUs associated with the highest reconstruction errors ($f_8$ and $f_{19}$) and the lowest reconstruction errors ($f_7$ and $f_{14}$) are shown in the figure.

\begin{figure}[htbp]
    \centering
    \includegraphics[width=1\linewidth]{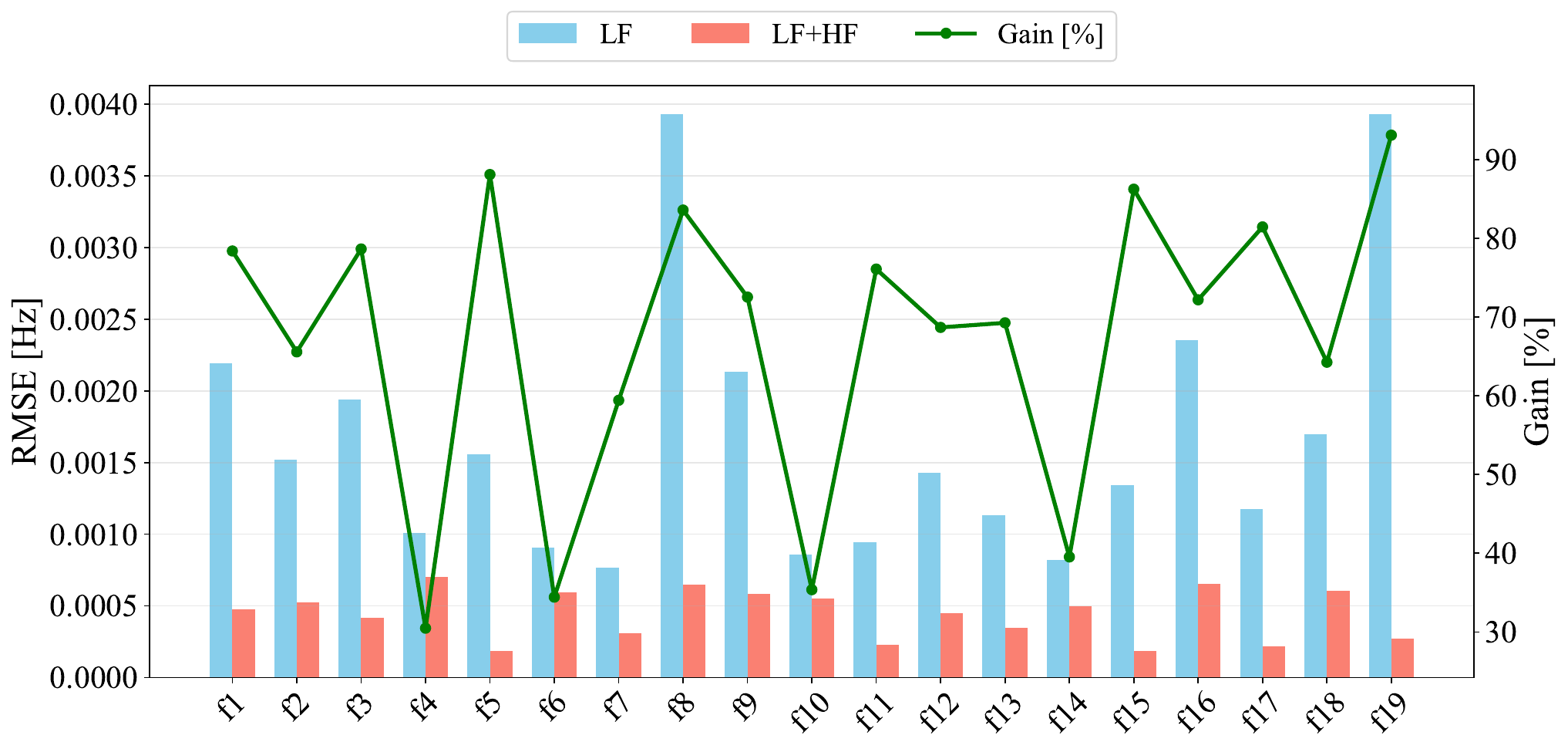}
    \caption{RMSE [Hz] and Gain [\%] between LF and LF+HF for all the PMUs reconstructions in March 2021.}
    \label{fig:RMSE_Gain_plot_2021}
\end{figure}

As can be inferred, the LF pathway is able to identify the dominant oscillatory mode (i.e., the interarea mode), while the inclusion of the HF pathway enables SINDy-SENDAI to accurately reconstruct the complete PMU measurements. To further quantify the reconstruction performance, Fig.~\ref{fig:RMSE_Gain_plot_2021} presents the RMSE values for both the LF and LF+HF pathways across all PMU measurements, as well as the corresponding gain.

\begin{figure*}[htbp]
    \centering
    \includegraphics[width=1\linewidth]{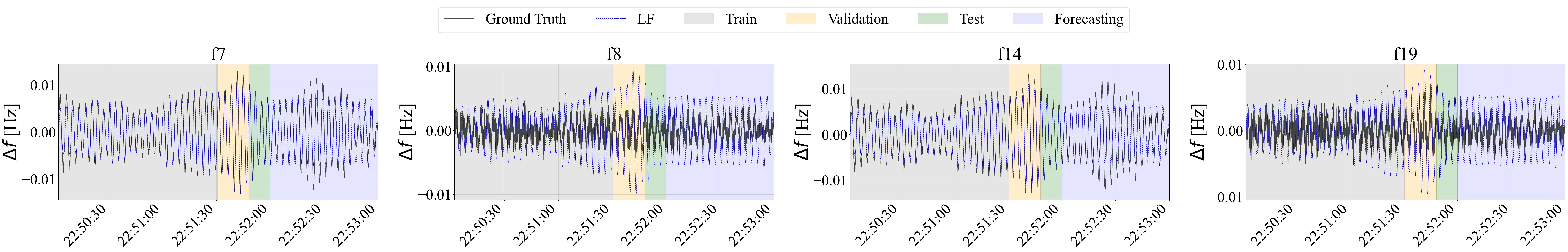}
    \caption{One-minute forecast of the LF pathway for $f_7$, $f_8$, $f_{14}$ and $f_{19}$ PMUs. The SINDy model accurately reproduces the system dynamics during the first seconds, after which errors gradually accumulate, leading to divergence from the true trajectory.}   
    \label{fig:PMU2021_inference_w1}
\end{figure*}

As observed, the largest LF reconstruction errors occur at sensors $f8$ and $f19$, located in the northern region of the Italian power system, where the dominant dynamics captured by the LF pathway are less pronounced. The integration of the HF pathway allows SINDy-SENDAI to compensate for these local discrepancies, significantly enhancing reconstruction accuracy across the entire sensor network.
The linear system identified in the train time window by SINDy is:
\begin{equation}\label{eq:sindy_ITA}
\resizebox{0.89\columnwidth}{!}{$
\begin{cases}
\dot{z}_1 = 0.440\, z_1 - 0.651\, z_2 + 0.332\, z_3 + 0.819\, z_4 + 0.032 \\
\dot{z}_2 = 2.216\, z_1 - 66.516\, z_2 + 61.457\, z_3 + 7.229\, z_4 - 0.190 \\
\dot{z}_3 = 2.221\, z_1 - 66.608\, z_2 + 61.531\, z_3 + 7.249\, z_4 - 0.188 \\
\dot{z}_4 = 1.058\, z_1 - 37.340\, z_2 + 33.747\, z_3 + 4.538\, z_4
\end{cases}
$}
\end{equation}
The identified linear mode can be integrated using standard ODE solvers to forecast the near-future evolution of the LF pathway. Figure~\ref{fig:PMU2021_inference_w1} shows the resulting forecasts over a one-minute horizon, extending up to 22:53:00.
As observed, the SINDy model accurately reproduces the LF dynamics for several tens of seconds following training. 

Finally, the dynamic stability of the linear model identified in \eqref{eq:sindy_ITA} can be assessed via eigendecomposition analysis. The resulting estimates of modal frequency and damping ratio are reported in Table~\ref{tab:Iberian2021} and compared with those obtained using the hDMD in the full measurements space. 
\begin{table}[htbp]
\centering
\caption{Modal parameters estimation results in March 2021.}
\label{tab:Iberian2021}
\begin{tabular}{c|c|c}
\toprule
&\textbf{hDMD} & \textbf{SINDy-SENDAI} \\
\midrule
$f_1$ [Hz] & 0.286 & 0.277 \\
$\xi_1$ [\%] & 0.150 & 0.24 \\
\bottomrule
\end{tabular}
\end{table}
As can be seen, SINDy-SENDAI captures the dominant interarea mode, yielding comparable frequency and damping estimates while providing an interpretable latent-space model.

\section{Limitations and Future Work}\label{sec:limitations}

One of the main challenges for real-world deployment is the need for hyperparameter tuning, which may limit the applicability of the proposed approach in real-time environments. Future work will therefore focus on the development of automated tuning procedures to reduce operator intervention and improve adaptability across different operating conditions.

Another limitation of the current framework is its inability to distinguish forced oscillations from natural electromechanical modes. To address this issue, future research will investigate the integration of SINDy with control (SINDYc), allowing exogenous inputs to be incorporated into the latent-space representation. This extension is expected to facilitate the identification and characterization of forced oscillations, enabling their separation from natural system dynamics and supporting the localization of their forcing sources.

\section{Conclusions}\label{sec:conclusions}
This paper introduces SINDy-SENDAI, an efficient deep learning algorithm capable of decomposing and reconstructing both low- and high-frequency patterns from sparse measurements while simultaneously identifying interpretable governing equations in the latent space. The proposed approach is first validated in a simulation environment using the two-area Kundur system, where its modal identification performance is assessed against modal analysis results, which are used as ground truth. Subsequently, the method is tested on two real-world datasets: the Iberian power system oscillatory event of December 2016 and normal operating conditions in the southern part of the Italian power grid of March 2021. The proposed approach outperforms the hDMD method, which is currently regarded as a state-of-the-art technique among European TSOs. Moreover, it enables the identification of a linear SINDy model in the latent space, providing a compact representation suitable for dynamic stability assessment, mode shape analysis, and short-term forecasting of electromechanical oscillations.

Given the societal importance of modern power systems, SINDy-SENDAI represents many aspects of an algorithm that are needed for practical implementation: (i) robustness, (ii) interpretability, and (iii) stable performance.  This is demonstrated in simulation and in with real data.  And instead of a deep learning algorithm which produces a black-box model beyond the capacity of interpretability, the architecture proposed allows for the potential human understanding by framing the dynamics in the framework of dynamic equations of motion.

\section*{Code}
The data that has been used is confidential. The code is available at: \href{https://github.com/Andrea-Pomarico/SINDy-SENDAI}{\textcolor{blue}{https://github.com/Andrea-Pomarico/SINDy-SENDAI}}.

\section*{Acknowledgments}
The authors acknowledges support from the Air Force Office of Scientific Research  (FA9550-24-1-0141).

\bibliography{biblio.bib}
\bibliographystyle{IEEEtran}

\newpage

\vfill

\end{document}